\documentstyle[11pt,paspconf]{article}

\begin{document}
\title{VLA 8.4 GHz monitoring observations of the CLASS gravitational lens B1933+503}
\author{I.W.A. Browne, E. Xanthopoulos, A.D. Biggs, M. Norbury}
\affil{University of Manchester, NRAL Jodrell Bank, Macclesfield, Cheshire SK11 9DL, England}

\begin{abstract}
The complex 10-component gravitational lens B1933+503 has been monitored with the 
VLA during the period February to June 1998. Here we present the results of 
this A-configuration 8.4 GHz monitoring campaign. The 37 epochs of observations have an
average spacing of 3 days. The data have yielded light curves for the four flat-spectrum 
radio components (components 1, 3, 4 and 6). 
We find no significant flux density changes in any of the four flat spectrum components. 
\end{abstract}

\keywords{observations, gravitational lensing, individual: B1933+503}

\section{Introduction}
B1933+503 is the most complex arcsec-scale gravitational lens system known and was  
found in the Cosmic Lens All-Sky Survey (CLASS; Browne et al. 1998).
Sykes et al. (1998) report the discovery of this lens and present VLA and MERLIN maps
that reveal up to 10 components, four of which are compact and have flat spectra while the rest
are more extended and have steep spectra. 
Components 1, 3, 4 and 6 are believed to be the lensed images of the flat spectrum core (Figure 1) and  
the rest of the components the lensed images of two compact 
``lobes" symmetrically situated on opposite sides of the core. 
An HST/WFPC2 image of B1933+503 (Sykes et al. 1998) shows a faint galaxy
with a compact core (the lensing galaxy). 
The redshift for the lensing galaxy is 0.755 and for the source 2.62 (Norbury et al. 1999 in preparation). 
Sykes et al. (1998) report variability of  
as much as 33\% at 15 GHz over the timescale of a couple of months and Nair (1998) has estimated 
through modeling of the system, time delays of 8, 7 and 9 days for components 3, 4 and 6 with respect to 1. 
We report results of a four month monitoring campaign with the VLA at 8.4 GHz in its A-configuration.

\begin{figure}
\plotfiddle{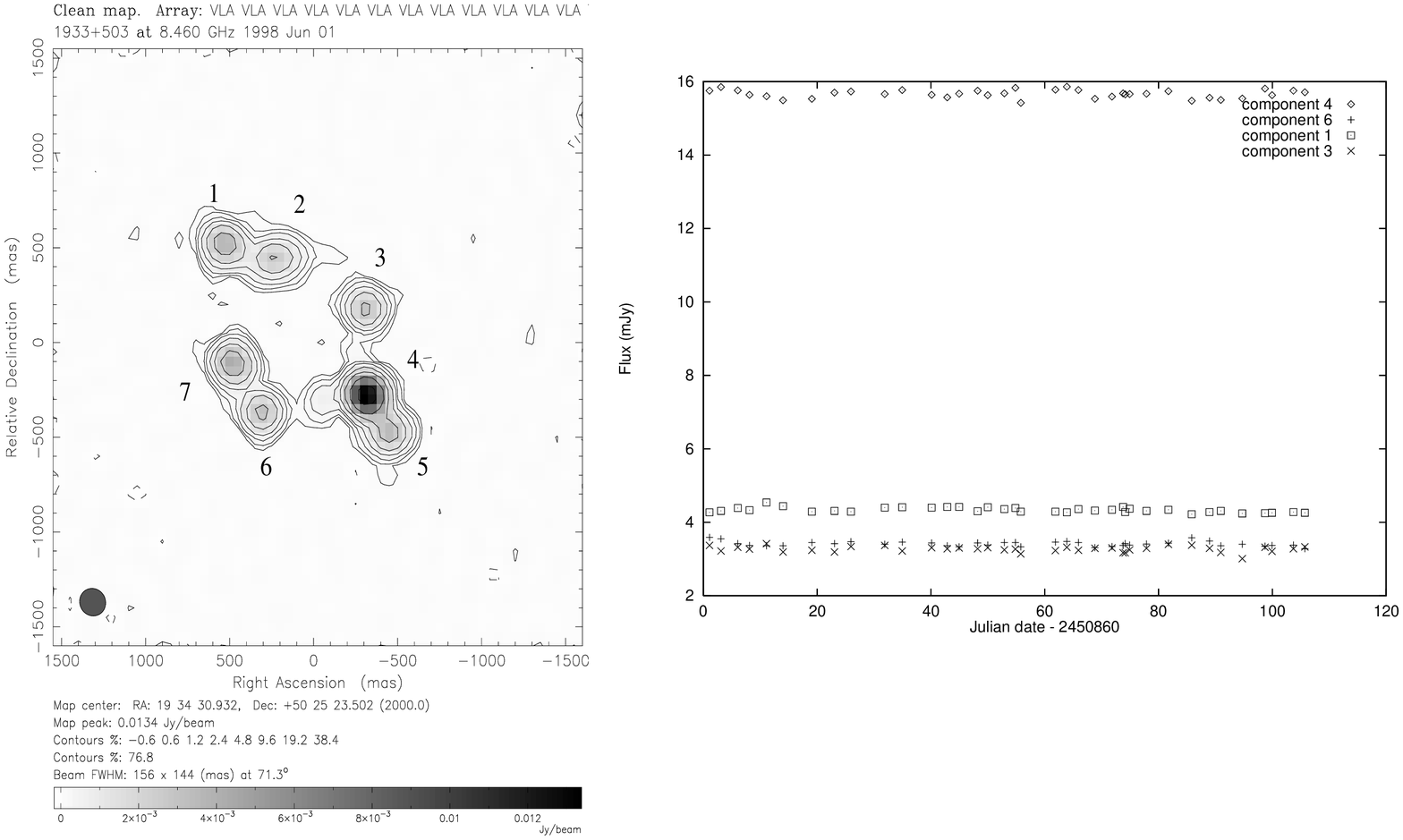}{6.5cm}{0}{40}{40}{-140}{-10
}
\caption{Left: The VLA map from the 37 epochs of observations Right: The light curves of components 1, 3, 4 and 6 of B1933+503}
\end{figure}

\section{Observations and Reductions}

B1933+503 was observed between February 16 1998 and June 1 1998. The resolution is  
240 mas. 
The 37 epochs were separated, on average, by 3 days. 
The observing strategy was kept consistent over the period of the observations. Each epoch contains two or
three 6 min observations of B1933+503 sandwiched between two observations of the steep spectrum source
1943+546 
of approximately 1 min each that was used for amplitude
and phase calibration purposes. A typical epoch has 30 min observing time.
Calibration was performed using the NRAO Astronomical Image Processing Software package AIPS. 
Once calibrated, we use the package DIFMAP to map and model fit the
UV-data.  
\section{Conclusions}
Figure 1 shows the results of the modelfitting (the light curves) to the 4 flat spectrum components 1, 3, 4 and 6.
Any change in the flux density is less than 1.6, 2.4, 0.7 and 2.0 \% for  1, 3, 4 and 6 respectively.
Thus our first monitoring campaign of B1933+503 has shown no signs of variability of this system.
We believe that we have caught this lens system at its quiescent phase.  
We started in June 1999, a new 
monitoring campaign of this system with the hope that it might behave like  
B1600+434 which showed variability only during the second monitoring campaign (Koopmans et al. 1999). 
The unique complex structure of B1933+503 would make it then one of the most exciting systems to measure
the Hubble constant.
\acknowledgments
This research was supported by European Commission, TMR Programme, Research Network Contract ERBFMRXCT96-0034 \\ 
``CERES".


\begin{references}
\reference Browne, I. W. A. et al. 1998, in Observational Cosmology with the new radio surveys, 
Astrophysics and Space Science Library, Vol. 226 
\reference Koopmans, L. V. E. et al. 1999, \astap, submitted
\reference Nair, S. 1998, \mnras, 301, 315 
\reference Sykes, C. M. et. al. 1998, \mnras, 301, 310 
\end{references}
\end{document}